\documentclass[preprint,tighten]{aastex62}

\usepackage[utf8]{inputenc}
\usepackage{natbib}
\usepackage{graphicx}

\begin{document}

\newcommand{\pg}{PGIR 20dci}
\newcommand{\pgitself}{PGIR}

\journalinfo{accepted to AJ on 5 February 2021}

\title{Outbursting Young Stellar Object PGIR 20dci in the Perseus Arm}
\author{Lynne A. Hillenbrand}
\affiliation{Department of Astronomy, MC 249-17, California Institute of Technology, Pasadena, CA 91125, USA}
\email{lah@astro.caltech.edu}

\author[0000-0002-8989-0542]{Kishalay De}
\affiliation{Department of Astronomy, MC 249-17, California Institute of Technology, Pasadena, CA 91125, USA}

\author[0000-0001-9315-8437]{Matthew Hankins}
\affiliation{Arkansas Tech University, Russellville, AR 72801, USA}

\author[0000-0002-5619-4938]{Mansi M. Kasliwal}
\affiliation{Department of Astronomy, MC 249-17, California Institute of Technology, Pasadena, CA 91125, USA}

\author[0000-0001-6381-515X]{Luisa M. Rebull}
\affiliation{IPAC, California Institute of Technology, Pasadena, CA 91125, USA}

\author{Ryan M. Lau}
\affil{Institute of Space \& Astronautical Science, Japan Aerospace Exploration Agency, 3-1-1 Yoshinodai, Chuo-ku,
Sagamihara, Kanagawa 252-5210, Japan}

\author[0000-0002-0077-2305]{Roc M. Cutri}
\affiliation{IPAC, California Institute of Technology, Pasadena CA 91125}

\author[0000-0003-1412-2028]{Michael C.~B.\ Ashley}
\affiliation{School of Physics, University of New South Wales, 2052, Australia}

\author[0000-0003-2758-159X]{Viraj R. Karambelkar}
\affiliation{Department of Astronomy, MC 249-17, California Institute of Technology, Pasadena, CA 91125, USA}

\author{Anna M. Moore}
\affiliation{Australian Nationa1 University, Research School of Astronomy and Astrophysics, Mount Stromlo Observatory, Cotter Road, Weston Creek 2611 Australia}

\author[0000-0001-9304-6718]{T. Travouillon}
\affiliation{Australian Nationa1 University, Research School of Astronomy and Astrophysics, Mount Stromlo Observatory, Cotter Road, Weston Creek 2611 Australia}

\author{A.~K.~Mainzer} 
\affil{University of Arizona, Department of Planetary Science, Tucson, AZ 85721}

\begin{abstract}
We report the discovery of a likely outbursting Class I
young stellar object, associated with the star-forming region NGC 281-W
(distance $\sim 2.8$ kpc).  The source is currently seen only
at infrared wavelengths, appearing in both the 
Palomar Gattini InfraRed ($1.2~\mu$m) and the 
Near Earth Object Widefield Infrared Survey Explorer ($3.4$ and $4.6~\mu$m) 
photometric time-domain surveys.  
Recent near-infrared imaging reveals a new, extended scattered light nebula.
Recent near-infrared spectroscopy confirms the similarity of PGIR 20dci
to FU Ori type sources, based on strong molecular absorption in CO, H$_2$O, and OH, 
weak absorption in several atomic lines, and a warm wind/outflow as indicated by 
a P Cygni profile in the \ion{He}{1} 10830 \AA\ line.
This is a rare case of an FU Ori star with a well-measured long term
photometric rise before a sharper outburst, and the second instance 
of an FU Ori star with a documented two-step brightening in the mid-infrared.
\end{abstract}

\section{Introduction}

Episodic accretion in young stellar objects is an important aspect
of mass accumulation in stars. It also plays a determining role 
in the evolution of circumstellar disks, and likely influences 
the process of planet formation, as well as planetary survival.
\cite{armitage2015} provides a review of disk physics in young stars,
including nonsteady accretion and disk instabilities.  

In ``classical" T Tauri type young stars, the accretion process is 
controlled mainly by physics in the inner disk and
magnetospheric region that connects the disk to the star.  
Observed time-variability in broad-band photometry and in emission line profiles
of these sources is dominated by stochastic behavior, 
with timescales from hours to days \citep[e.g.][]{findeisen2015,cody2018,costigan2014}.
The empirical zoo of accretion-related photometric behavior
has been measured in unprecedented detail for young stars
by space missions including CoRoT, Spitzer, Kepler, and now TESS, 
all of which sample timescales from minutes to a few months.
Ground-based surveys such as 
ASAS \citep{shappee2014}, 
PTF \citep{law2009}, 
ZTF \citep{bellm2019,graham2019}, and 
ATLAS \citep{tonry2018} 
in the optical,
and VVV \citep{minniti2010} 
in the infrared, by contrast, 
typically have lower precision and cadence (day to a few days), 
but the advantage of long duration monitoring.  
Gaia and WISE$+$NEOWISE also provide precise space-based photometric data, 
but at much lower cadence (months), with total duration exceeding a decade now.
Many of these surveys have all-hemisphere or even all-sky coverage, 
enabling rare event detection -- including episodic accretion outbursts 
in very young stars.

Accretion-driven outbursts associated with young stellar objects 
have been identified at an increasing rate over the past decade.
The discoveries have come from both the ground-based and the spaced-based platforms.
Infrared surveys have the advantage of being able to penetrate the
high levels of extinction towards the molecular clouds where
young stars are born, and to some extent the additional circumstellar
extinction of the youngest self-embedded sources.

Although young star outbursts appear to come in a continuum of amplitudes
and rise times, for historical reasons \citep{herbig1977},
the labels EX Lup type bursts (also known as EXors)
and FU Ori type outbursts (also known as FUors) are often applied.
The former are more frequent, characterized by brightness increases of a 
few magnitudes in amplitude, lasting a few months
to one-to-two years, and have emission line spectra.  
The latter are more infrequent, and can be up to 5-6 mag 
in amplitude, lasting decades to centuries, and with absorption line spectra
except for P Cygni structure in some wind lines.
An increasing number of objects are unable to be cast neatly into
either of these categories, however.
Our knowledge of the spectrum of amplitudes, durations, and
duty cycles for young star outbursts is still in its infancy.  

The large-scale FU Ori outburst events are intrinisically rare. 
Furthermore, those few that have been captured in-progress
have not been particular well-characterized\footnote{The situation can be
compared to shorter timescale rare occurences such as tidal disruption events
and nearby supernovae, which garner armies of dedicated followers.}.
The most recent FU Ori outburst discoveries are improving the situation, e.g.
Gaia 17bpi \citep{hillenbrand2018}, Gaia 18dvy \citep{se2020}, 
and NWISE-F J213723.5+665145 \citep{stecklum2020,connelley2020}. 
The total census of FU Ori type stars numbers $<30$ objects at present, 
with less than half of these having had their outburst actually observed 
-- rather than inferred from imaging at well-separated pre-outburst and post-outburst epochs.

We report here on the discovery of substantial photometric brightening
of an object designated as \pg. This source is located at 
R.A. = 00:52:20.21, Dec. = $+$56:34:03.9,
and we associate it with the NGC 281-W star forming region.
The object appears to be a bona fide FU Ori star that has just reached
a plateau phase in its brightening.

\section{Source Environment}

NGC 281, also known as Sh 2-184, is an \ion{H}{2} region located in the Perseus spiral arm.
While the traditional distance in the literature on the stellar population is $\sim$2.2 kpc,
astrometry from Gaia DR2 \citep{gdr2} indicates a somewhat further 2.8 kpc.
This is also the distance derived by \cite{sato2008} 
from parallax studies at radio wavelengths of a maser population 
associated with NGC 281-W (where \pg\ is located).  These results place the region 
on the far side of the Perseus arm, and about 300 pc out of the Galactic plane.

IC 1590 ($<$2 Myr) is the main stellar cluster within the nebular region, 
and contains the bright ionizing sources collectively designated as HD 5005.
This source is now a spatially resolved multiple star system, 
consisting of components with spectral types O6.5, O8, and O9 \citep{guetter1997}. 
The cluster thus draws analogies in the literature as a Trapezium type system. 
The stellar population of the IC 1590 cluster was investigated by \cite{guetter1997} 
and by \cite{sharma2012}, the latter of whom find a size $r \approx 6.5$ pc.

To the southwest of the main cluster IC 1590, is a dark lane known as NGC 281-W.
Another dark area just to the east of the nebulosity is designated NGC 281-E.  
These are the locations of molecular clouds 
at the same distance as the large nebula, and presumably situated,
at least in part, on its near side.
APOD provides an optical picture\footnote{\url{https://apod.nasa.gov/apod/ap110825.html}}
that may clarify the geometry for some readers. 
CO maps of the overall region can be found in \cite{elmegreen1978} 
and \cite{lee2003}, who provide evidence for cloud compression coming from
the direction of the cluster and \ion{H}{2} region.

The NGC 281-W molecular core has a bright sub-millimeter and far-infrared source,
readily apparent in Herschel/SPIRE \citep{griffin2010} and SCUBA-2 \citep{holland2006} maps.  
This dust clump harbors  a collection of mid-infrared point sources 
to its east and southeast, as seen in WISE \citep{wright2010} images.

The portion of this highly embedded cluster that is visible in the near-infrared 
was discovered and studied by \cite{carpenter1993}. These authors targeted 
the region with early near-infrared imaging capabilities based on 
the position of IRAS 00494+5617.  They found a cluster size r = 0.74 pc 
(corrected here for differences in the distance assumption), 
with northeastern and southwestern portions separated 
by a ridge of high extinction.  The ``typical" extinction towards
the embedded sources was estimated at $A_V \approx 10$ mag, with peak $A_V \approx 45$ mag.
\cite{carpenter1993} also derived a mass of 340 $M_\odot$ 
(corrected here for differences in the distance assumption), 
for the associated 
CS core\footnote{A distance-corrected mass in CO of 4000 $M_\odot$ was also reported by
\cite{carpenter1993}, though the NGC 281-W cloud is somewhat larger than the 
area studied, with \cite{lee2003} finding a total mass for the extended cloud
of $1-3\times 10^4 M_\odot$.}.
The same area of NGC 281-W was specifically targeted in a similar study 
by \cite{megeath1997} using more sensitive radio and infrared observations. 

\begin{figure}
\begin{center}
\includegraphics[width=0.47\textwidth]{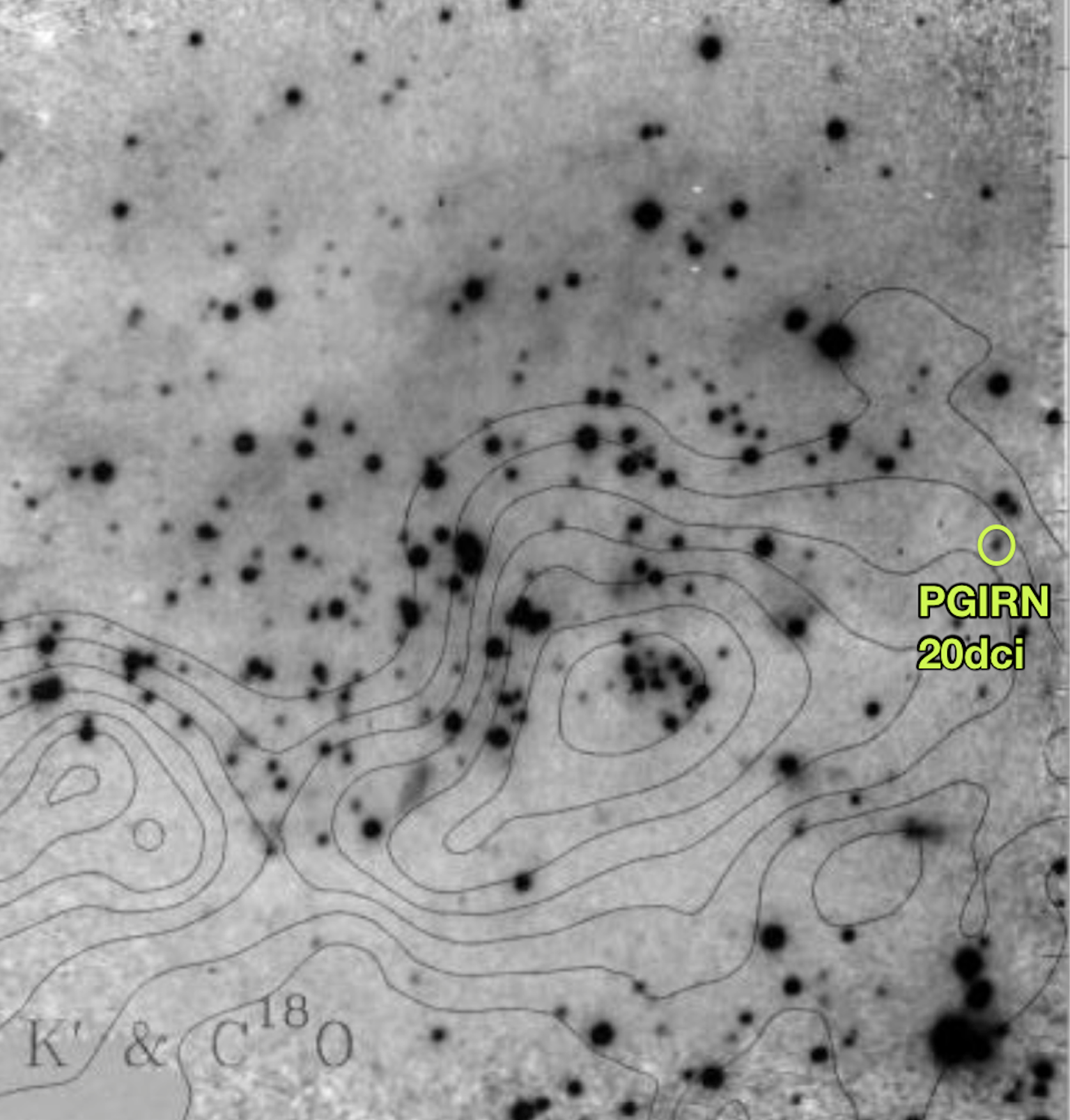}
\includegraphics[width=0.49\textwidth]{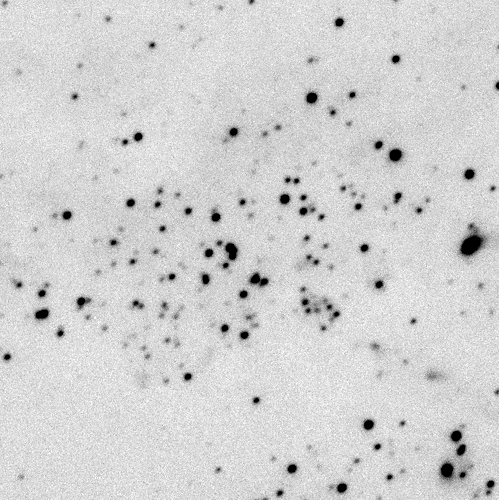}
\end{center}
\caption{
Left panel:
Figure reproduced from \citet[][their Figure 9]{megeath1997} 
showing a deep near-infrared image taken in $K'$ ($2.2~\mu$m) 
along with contours of the $C^{18}O$ molecular line emission intensity.
Faint tick marks along the right axis indicate 10\arcsec\ intervals,
with total field of view 2.\arcmin07 $\times$ 2.\arcmin07.
Overlaid in bright green is a marking for the position of \pg, 
which does appear as a faint source in this early 1990's image.  
Right panel: $Ks$ ($2.2~\mu$m) image obtained in 2020 September, showing a much brighter
and extended \pg, even though the image is shallower than the one on
the left. In both panels, the data are from infrared cameras
on the Palomar 200" telescope (PFIR = Prime Focus InfraRed, on the left, and WIRC on the right)
}
\label{fig:mw97}
\end{figure}

The object of interest here, \pg, is located within the region of these
previous surveys, about 30\arcsec\ west of the embedded cluster center.
Figure~\ref{fig:mw97} shows a portion of the field imaged by \cite{megeath1997}, 
with the position of \pg\ designated.
Although faintly visible at $K'$-band in this archival image, 
\pg\ has not been previously cataloged, studied, or characterized.

\section{Outburst Discovery and Validation}

In this section, we describe the initial photometric alert that drew our attention
to \pg.  We then detail the optical, near-infrared, and mid-infrared photometric forensics 
work leading to establishment of the outburst profile over time.   

\subsection{Gattini 1.2 $\mu$m ($J$-band) Lightcurve}

The Palomar Gattini InfraRed survey \citep[PGIR; ][]{de2020a,moore2019}.
observes the sky every two nights in the near-infrared $J$-band, 
to a median depth of J = 15.7 mag (AB).  The faint limit is much brighter
than this in crowded or confused regions, such as near the Galactic plane
and in high surface brightness regions, such as NGC 281-W.

On 2019, August 31, a ``hostless" source was first detected in \pgitself\
as a positive image subtraction, and flagged (by co-author KD) 
for photometric and spectroscopic follow-up.
The initial source position required some refinement due to the 
8\arcsec\ pixel size of \pgitself. There are no counterparts in optical 
\citep[e.g. PanSTARRS][]{flewelling2016}  
or near-infrared \citep[2MASS][]{skrutskie2006} survey catalogs, though a hard stretch
of the 2MASS images does reveal a faint detection at $Ks$ and perhaps $H$.  
A nearby WISE catalog \citep{cutri2013} source at 00:52:20.41 +56:34:03.3 
was determined to be the source that had brightened.
The brightening is also recorded in NEOWISE \citep{mainzer2014} data. 
The consequent increase in signal-to-noise likely yields better astrometry, 
and hence a more accurate true source position of 00:52:20.21 +56:34:03.9,
which we adopt.  This position is 1.3\arcsec\ away from the position reported 
for the fainter WISE source, and 4-5\arcsec\ from the \pgitself\ position. 

\begin{figure}
\begin{center}
\includegraphics[width=0.49\textwidth]{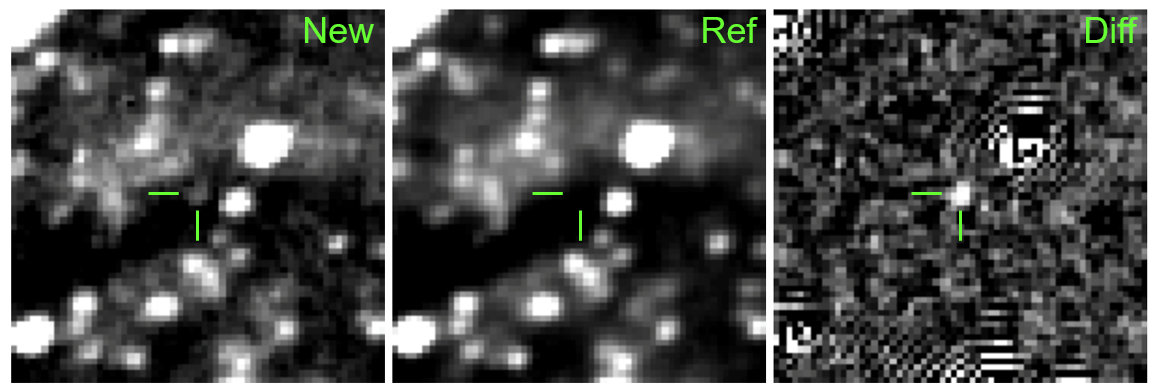}
\\
\includegraphics[width=0.55\textwidth]{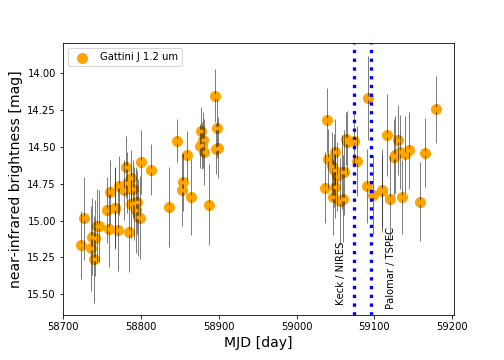}
\end{center}
\caption{
Top panels: Discovery imaging sequence showing 
a 4\arcmin $\times$ 4\arcmin\ region centered on \pg;
orientation is the standard north upward and east to the left.
From left to right are: the PGIR image from 2019-10-31 (MJD = 58787),
the static reference image of the field created 
from a series of earlier images taken between 2018-10-28 and 2019-01-29, 
and the subtracted difference image.
\pg\ is clearly visible in the last panel, and is marked in all three.
Bottom panel:
$J$-band lightcurve of \pg\ during 2019 and 2020.
The observed upward slope in the data points at MJD $<$ 59000
corresponds to a source brightening rate of $-0.116\pm0.011$ mag/month. 
The lightcurve then flattens, and a fit for MJD $>$ 59000
is consistent with zero slope ($-0.017\pm0.022$ mag/month).  
We note that $J$-band observations 
from much earlier indicate a significantly fainter source,
with estimated limits $J>17.1$ mag (2MASS) and $J>20$ mag \citep{megeath1997}; see text.
Vertical blue lines indicate the epoch of the infrared spectrum shown in Figure~\ref{fig:nires}
as well as a second spectral epoch discussed in the text. 
}
\label{fig:gattini}
\end{figure}

Figure~\ref{fig:gattini} shows the lightcurve produced from the \pgitself\
forced difference photometry pipeline. PSF-fit photometry is performed, with
noise uncertainty at the location of the transient estimated, as described 
in \citealt{de2020b}, from difference images 
produced using the ZOGY algorithm \citep{zackay2016}.

Since its discovery at $J=14.97$ mag, \pg\ has slowly increased in brightness.
The brightening trend observed during the 2019-2020 season (Figure~\ref{fig:gattini})
can be fit with a linear slope of $-0.116\pm0.011$ mag/month.
During the latter part of 2020 the lightcurve appears to have plateaued,
and a linear fit is consistent with zero slope.

\subsection{Pre-Discovery Photometric Information}

In this section, we discuss the limited photometric record available for \pg.

\subsubsection{Optical and Near-Infrared}

At the time of its \pgitself\ detection,
\pg\ was invisible in optical time-domain survey data, including ASAS and ZTF.  
Recently, however, a faint optical source with $r=20-20.5$ mag is transiently 
present in ZTF alerts \citep{masci2019}.  There is not a useful optical lightcurve, 
however, due to infrequent detection and large error bars.

We also note that \pg\ was undetected in deeper imaging 
of the NGC 281-W region with HST in the optical 
F814W and F675W filters, from 2000, September, 
accessible via the Hubble Legacy Archive.  
\pg\ is also undetected in archival Keck
images with DEIMOS of the region from 2010, October 6,
accessible via the Keck Observatory Archive.  

In the near-infrared, \pg\ was undetected by 2MASS \citep{cutri2003},
as noted above.  While the nominal 2MASS survey limit is $J < 15.8$, 
we derive a more accurate number for this exact
environment by considering the error vs magnitude distribution of sources 
within 4\arcmin\ of \pg. We find that $10\sigma$ photometry is available to $J=16.1$ mag
and $5\sigma$ photometry to $J=17.1$ mag. While the diffuse background 
is high in the NGC 281-W region, the dark cloud renders the point source background
and confusion lower than would be the case otherwise. 
Thus, the 2MASS magnitude limit is plausibly deeper than nominal.  
We conclude that a reasonable flux upper limit for \pg\ at epoch 2000, January 13
is approximately $J > 17.1 $ mag.  

In the deeper $J$-band imaging of \cite{megeath1997}, \pg\ is also not apparent, 
except for perhaps a faint low surface brightness smudge near the edge of detection.
The claimed faint point source limit of these data is $J\approx21$ mag. 
\pg\ does, however, have an apparent faint counterpart in the
K'-band imaging of \cite{megeath1997}, as indicated in Figure~\ref{fig:mw97}.
Although the sources from this study are not cataloged,
the authors report that the fainter stars in the displayed image 
have $K'\approx 18.5$ mag (80\% completeness level).
\pg\ thus appears to have experienced a $\sim$5 mag brightening since the 
mid-1990's epoch of the \cite{megeath1997} data acquisition.

\subsubsection{Mid-Infrared Spitzer}

In the mid-infrared, there is also a faint counterpart to \pg\ in Spitzer/IRAC
\citep{werner2004,fazio2004} imaging.  The source appears blue in comparison 
to the main embedded cluster, approximately 30\arcsec\ to the east of \pg, 
perhaps due to somewhat lower extinction away from
the center of the molecular core (see contours in Figure~\ref{fig:mw97}).

There are three relevant Spitzer/IRAC AORs, one for P.I.
Fazio in 2004 (AORKEY 4127744) and two for P.I. Wolk in 2009 (AORKEYs
34780160 and 34780416). The former, from Spitzer's cryogenic phase,
includes all four IRAC bands (3.6, 4.5, 5.8, and 8 $\mu$m).
The latter two were taken in the post-cryo phase of Spitzer and thus include 
only IRAC1 (3.6 $\mu$m) and IRAC2 (4.5 $\mu$m). We performed aperture
photometry at the target's location in the pipeline-produced post-BCD
mosaics using a radial aperture of 3 native pixels (1.2$\arcsec$/pixel) 
and a background annulus from 3-7 native pixels. Aperture corrections were 
applied as appropriate to the cryo and the post-cryo observations, 
as described in the IRAC Instrument
Handbook\footnote{\url{https://irsa.ipac.caltech.edu/data/SPITZER/docs/irac/iracinstrumenthandbook}.
Specifically, we used values of 1.124, 1.127, 1.143, \& 1.234 for the 4 channels in the cryogenic
era, respectively, and 1.125 \& 1.120 for the 2 channels in the post-cryo era, respectively.}. 
The Instrument Handbook also has the zero points\footnote{280.9, 179.7,
115.0, \& 64.13 Jy for the four IRAC channels, respectively.}
we used for converting the flux densities to magnitudes.
Our photometric results for Spitzer/IRAC are provided in the first three rows of Table~\ref{tab:phot}.

From some of these same Spitzer data, \cite{sharma2012} found [S3.6] or IRAC1 = 14.40 mag
and [S4.5] or IRAC2 = 12.89 mag using a similar 3.\arcsec6 radial aperture 
to us, but a smaller 3.\arcsec6--8.\arcsec4 sky annulus.  These values are
consistent with our results for the first observation epoch, in 2004.
We also note that IRSA's SEIP catalog\footnote{\url{http://irsa.ipac.caltech.edu/data/SPITZER/Enhanced/SEIP/overview.html}}
contains a source SSTSL2 J005220.20+563403.8 
with IRAC1 = 14.57 mag and IRAC2 = 12.80 mag, 
adopting the 3.\arcsec8 aperture values and applying the flux-to-magnitude conversions above,
again consistent with our measurements for the 2004 epoch, within the reported errors. 

Finally, we note that there are also Spitzer/MIPS observations of the region
available in the $24~\mu$m band.  These MIPS1 data were taken in 2006, February,
which is between the two Spitzer/IRAC epochs.   
It would be valuable to be able to compare the 2006 MIPS1 photometry 
to the 2010 W4 photometry, for reasons that will become apparent below. 
Unfortunately, however, the MIPS1 imaging results in only flux upper limits for \pg.

\subsubsection{Mid-Infrared AllWISE}

WISE images at the comparable bands WISE1 or $W1$ (3.4 $\mu$m) and WISE2 or $W2$ (4.6 $\mu$m) 
have lower spatial resolution, but do show a source  
that appears more distinct and pointlike in the unWISE image processing \citep{lang2014}.
The AllWISE catalog \citep{cutri2012} records a source WISEA J005220.41+563403.3 
with reported photometry given in Table~\ref{tab:phot}.
We took the AllWISE profile fitting magnitudes and errors, as quoted. 
The measurements are rated as ``AAAA" in photometric quality, meaning SNR $>10$ 
in each of the four bands, and they are unsaturated.

\begin{table}
\scriptsize
\begin{centering}
\caption{Spitzer, WISE, and NEOWISE Photometry
\footnote{IRAC columns contain Spitzer/IRAC photometry derived here, with transformations made to the WISE photometric system appearing in the WISE columns as parenthetical entries.
           WISE columns contain photometry from several sources.  
	   The first WISE entry is taken directly from the IRSA AllWISE catalog.  
	   Subsequent entries are derived here from observations reported in the IRSA MEP 
	   and NEOWISE-R catalogs, combining individual measurements  by ``visit"; see text.}
for PGIR 20dci}
\begin{tabular}{| l | c  c | c  c| c  c| c  c| c  c| c  c| c  c| c  c |}
\hline
MJD         & WISE1  & err& IRAC1& err&IRAC2& err& WISE2  & err&IRAC3& err&IRAC4& err&  WISE3& err& WISE4  &  err\\ 
            &3.4 $\mu$m & & 3.6 $\mu$m &    & 4.5 $\mu$m&    & 4.6 $\mu$m&    & 5.8 $\mu$m&    &8.0 $\mu$m &    & 12 $\mu$m&    & 22 $\mu$m &     \\
\hline
53214.65    &(15.01)&\nodata&14.49 &0.17&12.86&0.07&(12.61) &\nodata&11.77&0.30&11.56&1.60&\nodata &\nodata&\nodata &\nodata\\ 
55077.57    &(14.49) &\nodata&13.99 &0.12&12.42&0.07&(12.18) &\nodata&\nodata &\nodata&\nodata &\nodata&\nodata &\nodata&\nodata &\nodata\\
55080.44    &(14.45) &\nodata&13.96 &0.11&12.43&0.07&(12.19) &\nodata&\nodata &\nodata&\nodata &\nodata&\nodata &\nodata&\nodata &\nodata\\
\hline
(AllWISE)     &12.51&0.08& \nodata &\nodata&\nodata &\nodata&  11.40&0.05&\nodata &\nodata&\nodata &\nodata&6.11&0.04&3.36 &0.07  \\
55222.15    &12.63&0.25& \nodata &\nodata&\nodata &\nodata&  11.41&0.09&\nodata &\nodata&\nodata &\nodata&6.04&0.22&3.27&0.32\\
55409.82    &12.63&0.25& \nodata &\nodata&\nodata &\nodata&  11.41&0.09&\nodata &\nodata&\nodata &\nodata&6.13&0.22&3.37&0.32\\
55586.45    &12.62&0.25& \nodata &\nodata&\nodata &\nodata&  11.43&0.09&\nodata &\nodata&\nodata &\nodata&\nodata&\nodata&\nodata&\nodata \\
\hline
56684.21    &12.34&0.15&\nodata&\nodata&\nodata &\nodata&11.14&0.10&\nodata &\nodata&\nodata &\nodata&\nodata&\nodata&\nodata &\nodata  \\
56873.87    &12.56&0.22&\nodata&\nodata&\nodata &\nodata&11.14&0.03&\nodata &\nodata&\nodata &\nodata&\nodata&\nodata&\nodata &\nodata  \\
57045.96    &12.40&0.13&\nodata&\nodata&\nodata &\nodata&11.07&0.10&\nodata &\nodata&\nodata &\nodata&\nodata&\nodata&\nodata &\nodata  \\
57235.95    &12.43&0.21&\nodata&\nodata&\nodata &\nodata&11.12&0.07&\nodata &\nodata&\nodata &\nodata&\nodata&\nodata&\nodata &\nodata  \\
57408.47    &12.26&0.26&\nodata&\nodata&\nodata &\nodata&11.00&0.01&\nodata &\nodata&\nodata &\nodata&\nodata&\nodata&\nodata &\nodata  \\
57602.68    &12.31&0.26&\nodata&\nodata&\nodata &\nodata&10.97&0.07&\nodata &\nodata&\nodata &\nodata&\nodata&\nodata&\nodata &\nodata  \\
57769.55    &12.20&0.20&\nodata&\nodata&\nodata &\nodata&10.89&0.05&\nodata &\nodata&\nodata &\nodata&\nodata&\nodata&\nodata &\nodata  \\
57966.88    &12.32&0.17&\nodata&\nodata&\nodata &\nodata&10.85&0.06&\nodata &\nodata&\nodata &\nodata&\nodata&\nodata&\nodata &\nodata  \\
58129.81    &12.02&0.17&\nodata&\nodata&\nodata &\nodata&10.68&0.06&\nodata &\nodata&\nodata &\nodata&\nodata&\nodata&\nodata &\nodata  \\
58331.39    &12.09&0.18&\nodata&\nodata&\nodata &\nodata&10.62&0.05&\nodata &\nodata&\nodata &\nodata&\nodata&\nodata&\nodata &\nodata  \\
58491.38    &11.83&0.18&\nodata&\nodata&\nodata &\nodata&10.50&0.03&\nodata &\nodata&\nodata &\nodata&\nodata&\nodata&\nodata &\nodata  \\
58698.13    &11.17&0.10&\nodata&\nodata&\nodata &\nodata&9.70 &0.04&\nodata &\nodata&\nodata &\nodata&\nodata&\nodata&\nodata &\nodata  \\
58858.12   &10.35&0.09&\nodata&\nodata&\nodata &\nodata& 8.97&0.03&\nodata &\nodata&\nodata &\nodata&\nodata&\nodata&\nodata &\nodata  \\
59062.32   &10.25&0.10&\nodata&\nodata&\nodata &\nodata&8.89 &0.03&\nodata &\nodata&\nodata &\nodata&\nodata&\nodata&\nodata &\nodata  \\
59222.09   &10.20&0.08&\nodata&\nodata&\nodata &\nodata&8.93 &0.03&\nodata &\nodata&\nodata &\nodata&\nodata&\nodata&\nodata &\nodata  \\
\hline
\end{tabular}
\label{tab:phot}
\end{centering}
\end{table}

\subsubsection{Mid-Infrared Colors}

The available colors of the \pg\ progenitor are 
$IRAC1-IRAC2 = 1.60$ and $IRAC2-IRAC3= 1.09$ mag from Spitzer, and
$W1-W2 =  1.12$, $W2-W3 =  5.82$, and $W3-W4 =  2.76$ mag from AllWISE.

These colors place the object firmly within the Class I category 
of young stellar objects, meaning a steeply rising spectral energy distribution
that is best explained by the presence of a massive circumstellar envelope. 
The colors are too red to be consistent with a geometrically flatter disk, 
as is characteristic of Class II sources.  \pg\ is, however, even redder than 
most Class I sources, which indicates substantial foreground extinction.

\subsection{WISE$+$NEOWISE Lightcurve}

\begin{figure}
\begin{center}
\includegraphics[width=0.49\textwidth]{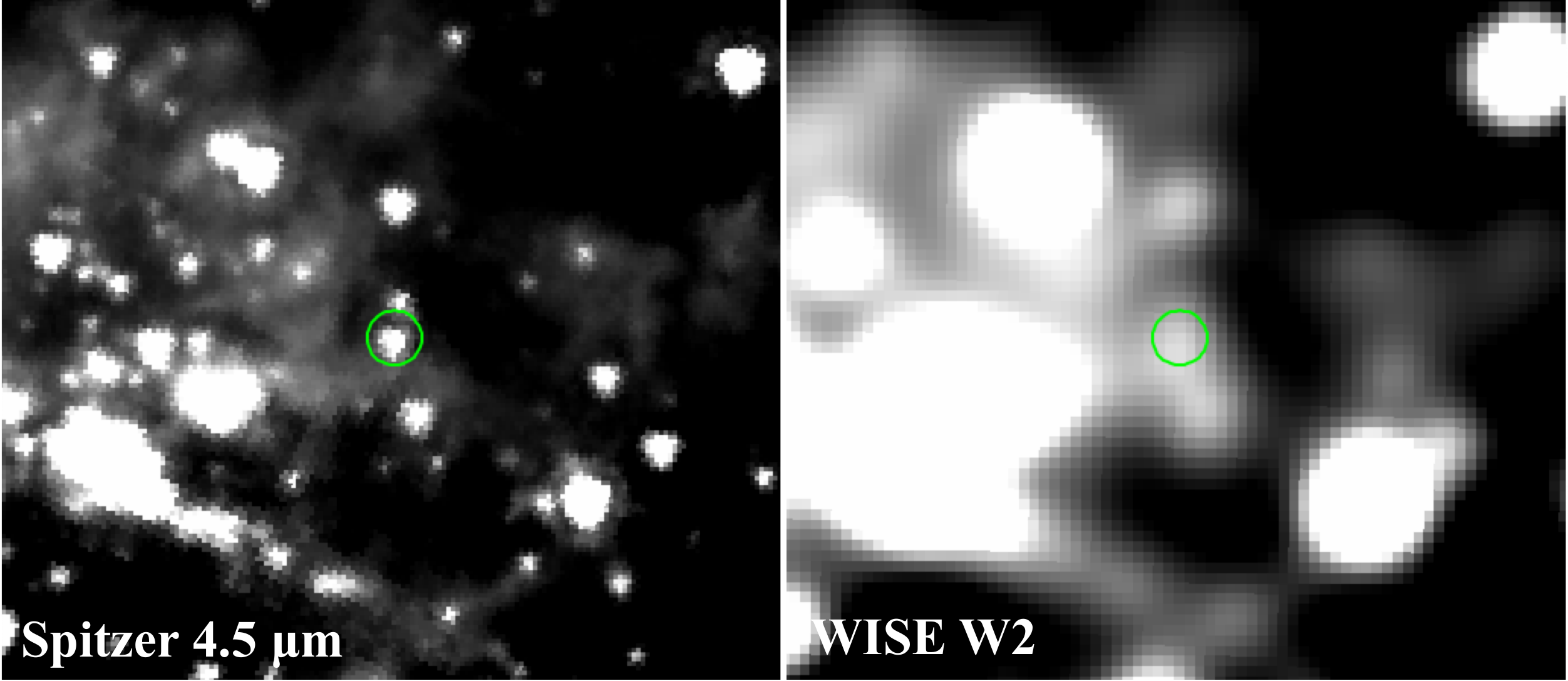}
\\
\includegraphics[width=0.55\textwidth]{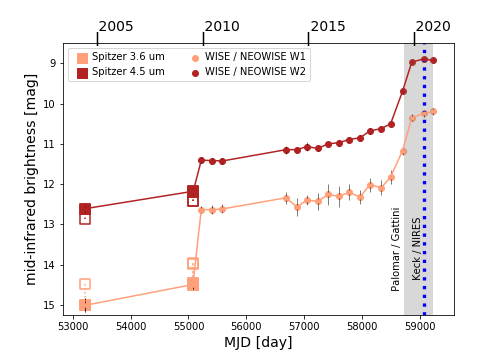}
\end{center}
\caption{
Top panels:
Spitzer 2009 and WISE 2010 images of a 103\arcsec\ $\times$93\arcsec\ field 
centered
on \pg; orientation is the standard north upward and east to the left.
As described in the text, Spitzer photometry was derived 
using an aperture radius of 3\arcsec6 (marked circle) while 
WISE photometry comes from the profile fitting measurements.
Although the WISE image is dominated by brighter sources, 
there are no confusion flags set in any of the WISE or NEOWISE 
catalog data for the profile fitting.
Bottom panel:
Mid-infrared lightcurve assembled from available 
Spitzer, WISE, and NEOWISE data. Error bars are indicated,
though in some cases are smaller than the data points.  
The Spitzer photometry is plotted both as observed (open squares)
and color corrected into the WISE photometric system (filled squares); see text.
Each of two large jumps in mid-infrared brightness is preceded by a shallow rise phase.
Gray bar marks the MJD coverage of the \pgitself\ lightcurve shown in Figure~\ref{fig:gattini}.
Vertical blue line indicates the epoch of the infrared spectrum shown in Figure~\ref{fig:nires}. 
}
\label{fig:midir}
\end{figure}

In this section, we present mid-infrared data that support the
\pgitself\ detection of a significant brightening of the source.

We collected photometry
from the WISE Multi-Epoch and NEOWISE \citep{cutri2015} mission archives
via IRSA\footnote{\url{https://irsa.ipac.caltech.edu/Missions/wise.html}},
rejecting measurements with qi\_fact=0 or qual\_frame=0. 
We checked the WISE/NEOWISE photometry flags\footnote{\url{https://wise2.ipac.caltech.edu/docs/release/neowise/expsup/sec2_1a.html}}, finding no entries
in the {\it w[12]cc\_map\_str} (contamination and confusion) column, and 
only non-event entries in the {\it na} (deblending) column.
The {\it nb} (multiple PSF) column indicates that a single profile is used
in $>80$\% of the measurements; only $<20$\% of the measurements
(all towards the fainter end) require a second PSF, perhaps to account for
the faint star just to the north of \pg\ (see the Spitzer image panel in Figure~\ref{fig:midir}).

For the WISE Multi-Epoch catalog and the subsequent, ongoing NEOWISE monitoring data, 
we took the median of all profile fitting magnitude measurements 
within each spacecraft ``visit" covering the position, and we calculated the error as the dispersion 
among these magnitudes.  The median time span per WISE or NEOWISE visit is about 1.25 days. 
Results are provided in Table~\ref{tab:phot}.

As illustrated in Figure~\ref{fig:midir}, the mid-infrared lightcurve of \pg\ 
experienced an exponential rise phase over the past few years, that appears to be plateauing during 2020.  
Considering only the three NEOWISE measurements at
MJD = 58491.8, 58698.2, and 58858.1, before the flattening,
the rise rate was $-0.121\pm 0.015$ mag/month or -1.40 mag/year during 2019-2020. 
We note that this is identical, within the errors, to the slope derived 
for the $J$-band lightcurve 
($-0.116\pm0.011$ mag/month; see discussion above regarding Figure~\ref{fig:gattini}).
Before this steep recent rise, there was a much shallower 
but steady upward trend in the WISE/NEOWISE lightcurve between 2010 and 2019, 
with a rise slope of only -0.01 mag/month or -0.13 mag/year.

\subsection{Overall Mid-Infrared Light and Color Evolution}

As discussed in 3.2.2, prior to the 2010 WISE mission measurements, 
there are data available in 2009 and 2004 from Spitzer. 
A shallow rise is also seen in these data, though there are a limited
number of data points.  The steeper jump that is implied between
the Spitzer and WISE epochs occurs in a fairly narrowly confined time frame, 
within 2009 September 6 (Spitzer) and 2010 January 26 (WISE), 
and warrants critical investigation.

The two shortest wavelength Spitzer and WISE filters 
are similar, and \cite{jarrett2011} find agreement in large-scale 
photometry comparisons to within 2-3\%.  However, \pg\ is an extremely red 
source, beyond the range of colors available to \cite{jarrett2011},
and thus we need to consider potential color terms between the
earlier Spitzer and later WISE photometry.  Similar considerations were
also made for the lightcurve analysis of Gaia 17bpi \citep{hillenbrand2018}
and PTF 14jg \citep{hillenbrand2019}, where color transformations were
derived\footnote{ 
$(W1 -W2)_{corr} = 1.62 \times (IRAC1-IRAC2) -  0.04$ mag, with rms = 0.24 mag.} 
and implemented, but not magnitude transformations.
Here for \pg, we have tried to assess more carefully the impact of color terms 
on the lightcurve, by looking at a large sample of young stellar objects 
observed with both Spitzer and WISE.  
While there is a narrow correlation in the magnitudes themselves, when considering
the IRAC1$-W1$ and IRAC2$-W2$ transformations as a function of IRAC1-IRAC2 color,
the comparison data show a fanning out towards redder IRAC1-IRAC2 color,
which introduces significant uncertainty.  
Identical issues were investigated by  \citet[][see their section 2.1.1]{anton2014}
and we have adopted their corrections\footnote{
$W1_{corr} = IRAC1 - 0.27489\times(IRAC2-IRAC1) + 0.07146$
and 
$W2_{corr} = IRAC2 + 0.1422\times(IRAC2-IRAC1) - 0.01855 $
}. 
For the IRAC1-IRAC2 color of \pg, this amounts to 
shifts of approximately 0.5 mag {\it fainter} in going from IRAC1 to $W1$, and -0.25 mag
brighter in going from IRAC2 to $W2$. Figure~\ref{fig:midir} illustrates both the
observed and the WISE-corrected Spitzer magnitudes, which also appear in the WISE columns of Table 1.

In order order to negate the conclusion
of a large jump in brightness between the last Spitzer epoch and the first WISE
epoch, the color corrections to the photometry would have to be much larger
than currently supposed (which is certainly possible), but also
{\it in the opposite direction} for the Spitzer IRAC1 channel.
We also note that, even among the very red sources in this spatial region,
\pg\ is an outlier in direct comparisons of Spitzer vs AllWISE photometry,
lending support to the reality of the apparent brightness jump. 

We can further scrutinize the WISE/NEOWISE photometry by considering
reported aperture magnitudes. The aperture photometry
exhibits the expected trends with aperture size, with the 5\arcsec.5 aperture
undersized and producing magnitude overestimates, and other apertures
oversized and thus having magnitudes that well-underestimate the profile fitting
magnitudes.  
Even if the undersized 5\arcsec.5 aperture photometry were used, the brightness jump
between the Spitzer 2009 and WISE 2010 measurements would remain, albeit 
reduced by about 0.25 and 0.75 mag in $W1$ and $W2$, respectively.
We note that the {\it w[12]flg\_1} (5.5\arcsec\ aperture confusion) column indicates
that about 8\% of the aperture photometry measurements have some identified 
source of confusion; this is not significant for the {\it w1mpro} magnitude
measurements that we have adopted, however.  
Overall, the profile fitting photometry is more robust 
to bad pixels, cosmic rays, and the need to establish aperture corrections.

Finally, we can consider the evidence for color changes as the source has brightened
in the mid-infared.  The recent steep brightening of \pg\ has been colorless, within the errors. 
The median color over the 2014-2020 NEOWISE epochs is consistent with a constant
value of $W1-W2 = 1.40$ mag and rms = 0.21 mag.  
Earlier, in 2010, however, the WISE colors were $W1-W2 \approx 1.23$ mag 
for each of the three epochs, so marginally bluer in the period just after
the 2009-2010 jump in brightness.  The even earlier 2009 and 2004 Spitzer colors 
are both IRAC1-IRAC2 $\approx 1.60$, 
which after applying the Spitzer-to-WISE color transformation quoted above,
results in a corresponding $W1-W2 = 2.55$ mag.  
Applying the \cite{anton2014} magnitude corrections detailed above results
in fairly similar colors of 2.40, 2.31, and 2.26 mag 
for the the three Spitzer epochs.  

The above suggests that the color of \pg\ 
became bluer by about 1 mag in $W1-W2$ as it brightened 
during late 2009 and very early 2010 period, before reddening 
by a small amount ($\approx0.2$ mag) and then continuing to brighten colorlessly.
It is hard to know how to interpret this overall behavior.  If driven by an accretion event,
one would expect some blueing behavior.  At the same time, there may be extinction
clearing going on, which would also imply a color change to the blue.
The neutral color behavior that is observed over the past five years as
\pg\ has brightened seems anomalous with respect to these expectations.

In summary, the mid-infrared lightcurve of \pg\ shows that the source
has undergone a long-timescale, two-step rise from
a faint state to a bright state over the past $\sim 15$ years.
During the period 2004-2019, the source increased its mid-infrared
brightness by $>5$ mag in total.  There was a shallow brightening
from 2004-2009, a first significant jump in brightness during a 4 month time frame
in late 2009 or early 2010, a second shallow rise phase, and a more recent
second significant brightness jump beginning in 2019 and plateauing in 2020.  

\section{Follow-up Observations and Data Analysis}

In this section, we describe the imaging and spectroscopy that we collected on \pg.

\subsection{Imaging}

\subsubsection{Data Acquisition}

\pg\ was flagged for follow-up with the Palomar 60" telescope and SEDM 
\citep[Spectral Energy Distribution Machine; ][]{blag2018} in its imaging mode.
A sequence in the $gri$ filters was acquired on 2020 August 14 (UT)
and reduced using the photometric reduction pipeline. 

On 2020 August 28, we obtained near-infrared imaging of \pg\
using the Wide Field Infrared Camera (WIRC; \citealt{wilson2003}) 
on the Palomar 200-inch telescope. The data were acquired 
as a series of dithered exposures in $J$, $H$, and $Ks$ bands 
for a total exposure time of 495, 330, and 300 seconds, respectively. 
The data were reduced, stacked and photometrically calibrated 
using the pipeline described in \cite{de2020a}.

\subsubsection{Findings}

\begin{figure}
\begin{center}
\includegraphics[width=0.49\textwidth]{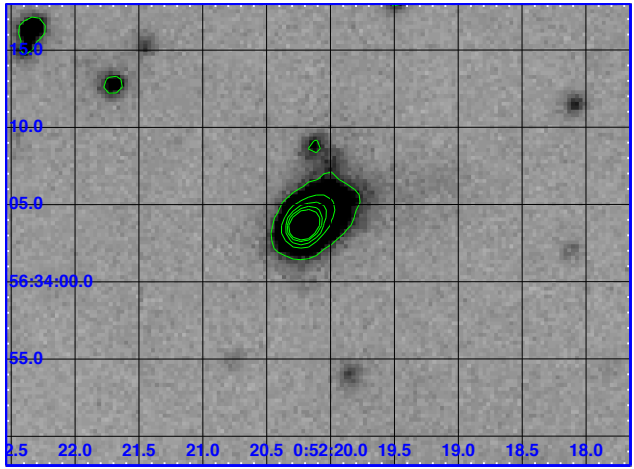}
\end{center}
\vskip-1truein
\caption{
Approximately 40\arcsec$\times$30\arcsec\ cutout from the
Palomar/WIRC $Ks$ (2.2~$\mu$m) image of Figure~\ref{fig:mw97}, centered near \pg.  
Grid spacing is 5 time seconds in R.A., and 5 arcseconds in Dec.
The extended structure is oriented southeast to northwest,
with concentrated flux towards the southeast and increasingly
diffuse emission towards the northwest, including beyond the lowest contour.
The extent of the scattered light within the contours is $\sim$14,000 AU.
}
\label{fig:guider}
\end{figure}

\begin{figure}
\begin{center}
\includegraphics[width=0.49\textwidth]{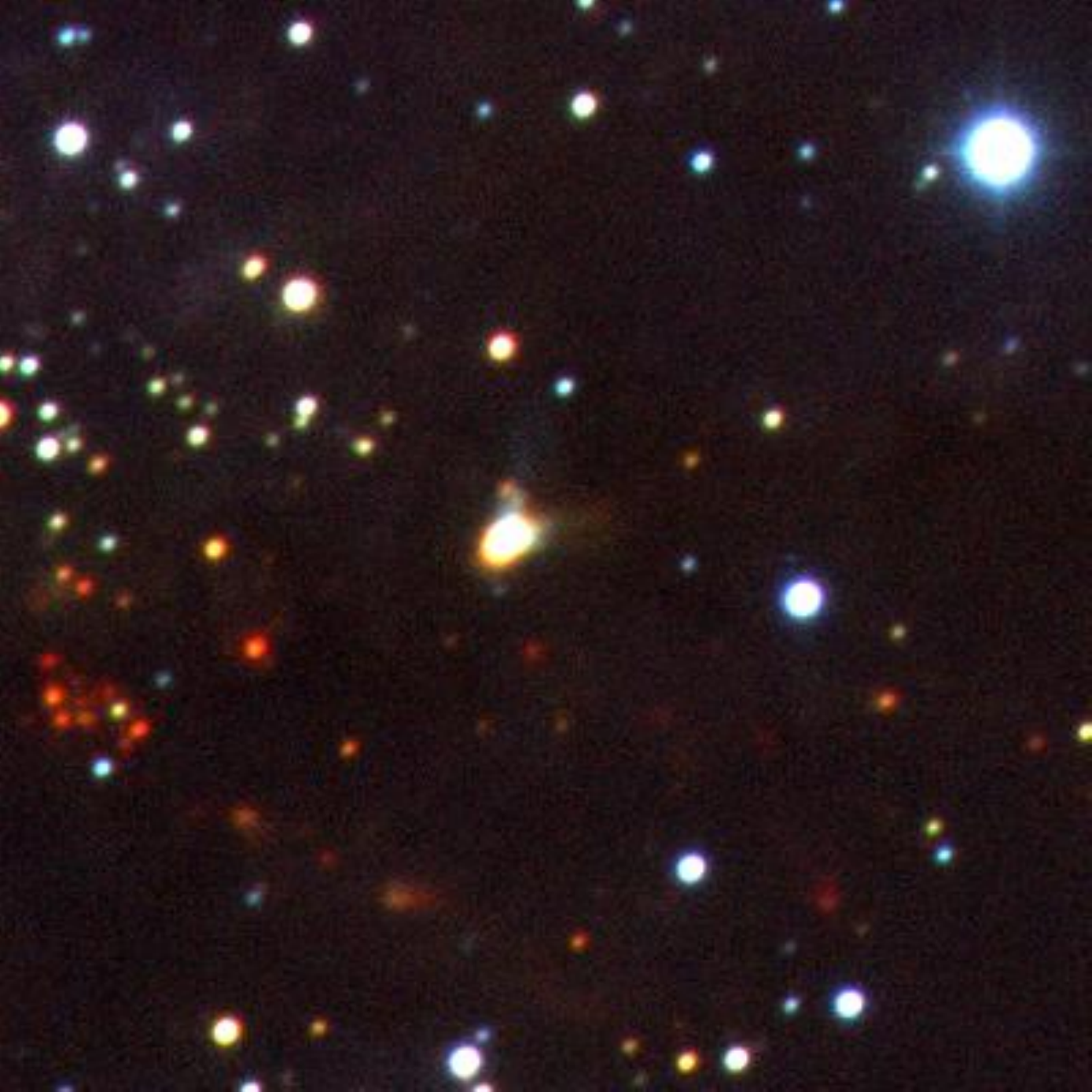}
\end{center}
\caption{
Color rendering of the Palomar/WIRC JHK data, in standard 
orientation of north upward and east to the left.
\pg\ in outburst is the extended source near center frame.
Faintly visible are extended ``tails" of emission to the northwest,
which may indicate the orientation of the outflow axis.
}
\label{fig:color}
\end{figure}

No counterpart to \pg\ was detected in $g$ and $r$ bands, 
with an approximate upper limit of $>21$ mag. 
However, we detect a faint extended source in $i$ band, 
coincident with the bright nebula detected in the near-infrared image. 
The final $Ks$-band image was shown in Figure~\ref{fig:mw97}.
A close-up on \pg\ appears in Figure~\ref{fig:guider}, which highlights
the extended cometary structure of the source in the near-infrared.
Figure~\ref{fig:color} shows the JHK color composite image.

Taken together, the optical and infrared imaging suggests 
an extremely red source, as well as a bright, extended, 
scattered light component associated with the brightening of \pg.  
The angular size of the extended structure corresponds to a physical size $\sim$14,000 AU.

In order to encompass the non-point source, extended nature of \pg, we
measured photometry from the Palomar/WIRC images using a fairly large aperture 
radius of 9 pixels (2\arcsec.25), whereas a 5-pixel (1\arcsec.25) radius
would suffice for point sources. Using the smaller radius would decrease 
the brightness of \pg\ by about 0.1 mag.  
We calibrated by matching to 110 sources 
in common between our final images and the 2MASS catalog.  
We find $J= 14.83$, $H=13.08$, and $Ks = 11.77$ mag in 2020, September.
The near-infrared colors of \pg\ in outburst are thus $J-H=1.75$ mag and $H-Ks=1.31$ mag.
This location in a $J-H$ vs $H-K$ diagram is consistent with that of a young stellar object
that is highly reddened.

\subsection{Spectroscopy}

\subsubsection{Data Acquisition}

\begin{figure}
\begin{center}
\includegraphics[width=0.99\textwidth]{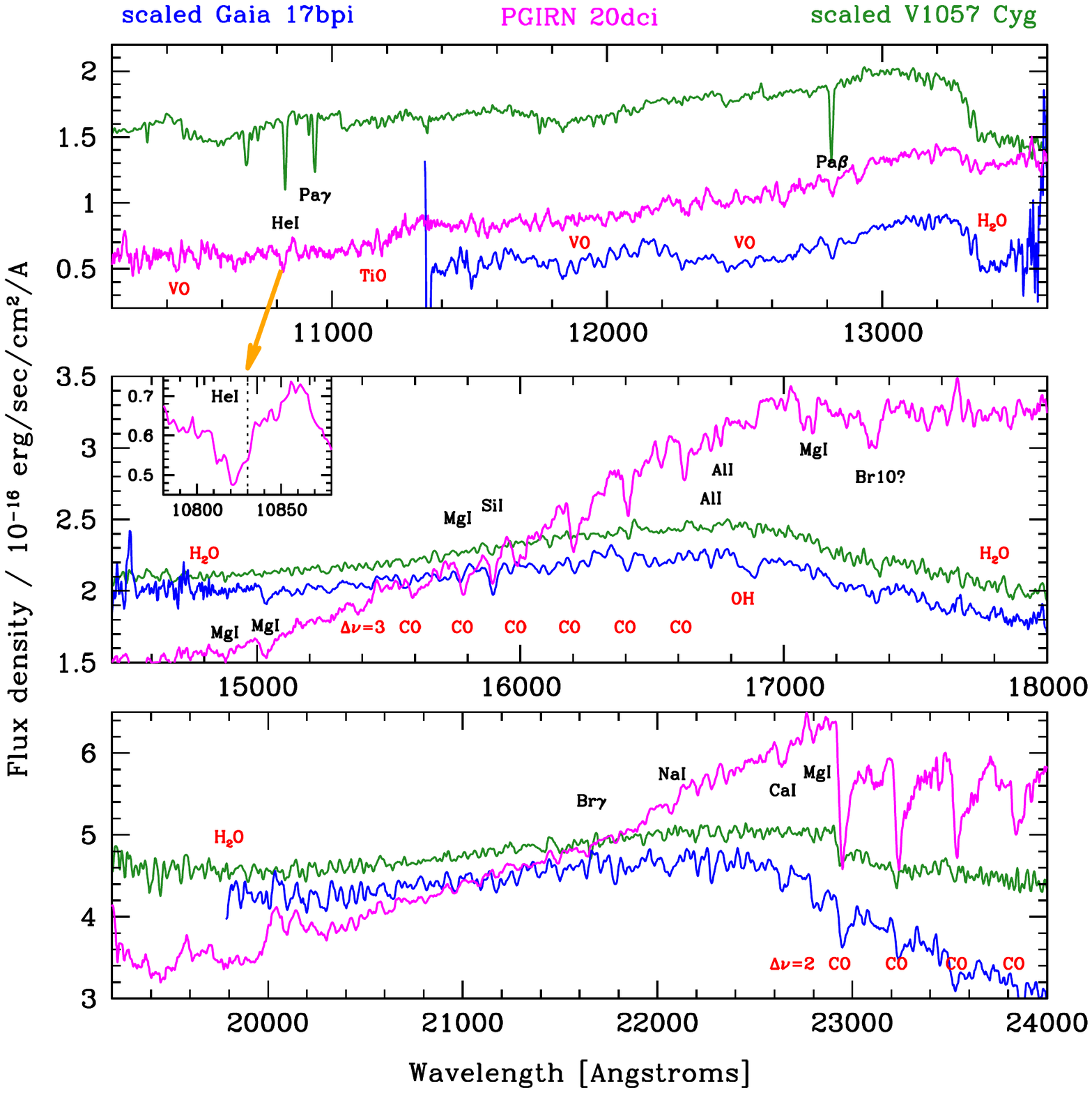}
\end{center}
\vskip-1.5truein
\caption{
Comparison of \pg\ (magenta) with the FU Ori stars 
V1057 Cyg \citep[green, from][]{connelley2018} and 
Gaia 17bpi \citep[blue, from][]{hillenbrand2018}. 
For purposes of plotting, the comparison objects 
have been scaled and shifted, with the corrections 
differing among the different panels.
The spectral match is apparent in molecular 
$^{12}$CO, H$_2$O, and OH, and possibly weak TiO and VO. 
Also present are atomic lines of
\ion{H}{1} Pa$\beta$ (and possibly Br10),
\ion{He}{1} 10830 \AA, 
and metals including \ion{Na}{1}, \ion{Mg}{1}, \ion{Al}{1},
\ion{Si}{1}, and \ion{Ca}{1}.
While the hydrogen and helium lines likely have some
outflow aspect to their absorption, the metal lines originate
in the photosphore, and combined with the molecular line absorption
indicate a mixed temperature spectrum.
}
\label{fig:nires}
\end{figure}

On 2020 August 13, \pg\ was observed with Keck II and NIRES
(Near Infrared Echellette Spectrometer), which is part of the family
of similar instruments described in \cite{wilson2004}.  
The detector records a prism-dispersed simultaneous $YJHK$ spectrum 
at resolution $R\approx$ 2700.  Four spectra were acquired using a 
6\arcsec\ ABBA nodding pattern, with integration time of 90 seconds at each position. 
The data were reduced using a modified version of \texttt{spextool} \citep{cushing2004}, 
and flux-calibrated with the telluric A0V standard star HIP 6002 
using the \texttt{xtellcor} code \citep{vacca2003}.
Figure~\ref{fig:nires} shows the final extracted and combined spectrum.
Due to the very red nature of the source, there is little signal at
the shorter wavelengths, and thus we show only the $JHK$ spectral region.

On 2020 September 4, a second spectrum was obtained at $R\approx 2700$
with the Palomar 200" and the TripleSpec \citep{herter2008} near-infrared spectograph.
Although lower signal-to-noise, and not shown as a result,
the spectrum is broadly the same in its shape and presence of strong spectral features
as the Keck/NIRES spectrum from a month earlier.  However, the flux level is brighter.
Considering Figure~\ref{fig:gattini}, the flux at this point in time could have been 
several tenths of a magnitude brighter, though the error on this estimate is large.
A more likely explanation is that the TripleSpec slit (1.0\arcsec) is wider than 
the NIRES slit (0.55\arcsec), and thus the spectrum would include more of the
bright extended structure of the source (see Figure~\ref{fig:guider}).
Normalizing the two at $K$-band, the $K$-band slope is identical to that shown in Figure~\ref{fig:nires},
but there is a sharper break just before the CO region, resulting in even deeper CO absorption.   
The $H$-band region is then slightly brighter than the spectrum shown in Figure~\ref{fig:nires}. 
This could mean that the source became slightly bluer while temporarily brighter at this epoch.  
However, we attribute the difference to the possibility
that the two spectra may have included different parts of the extended nebula.

\subsubsection{Findings}

Focussing on the Keck/NIRES spectrum, \pg\ is highly reddened, 
consistent with the extremely red mid-infrared colors measured by WISE.
The comparison objects\footnote{Gaia 17bpi is a recent FU Ori outburst 
occurring in 2017 \citep{hillenbrand2018}, which is still near its peak brightness. 
V1057 Cyg is one of the originally defined FU Ori stars, 
estimated to have outburst in 1970 \citep{herbig1977}, 
and although it has faded considerably, the spectrum maintains the general
features of FU Ori stars \citep[e.g.][]{connelley2018}.}
in Figure~\ref{fig:nires} 
are less reddened than \pg.  We can estimate the relative reddening, 
and thus the extinction, by applying a standard extinction law
to find the the best spectral match to the V1057 spectrum.
We find a differential extinction $\Delta A_V = 16$ mag, 
which can be added to a baseline $A_V$=4 mag for V1057 \citep{connelley2018}.
The total extinction towards \pg\ is thus estimated to be $A_V \approx 20$ mag.

Spectroscopically,
the most salient feature of \pg\ is the strong CO absorption.  
The source has both the $\Delta\nu = 2$ CO bands in the $K$-band,
and the $\Delta\nu = 3$ bands CO in the $H$-band.  
There is also strong H$_2$O in $J$-band and $H$-band, probably in $K$-band as well,
and likely $OH$ in $H$-band.  
Less obvious is the expected TiO and VO absorption, though it may be weakly present.
\pg\ shares with Gaia 17bpi and V1057 Cyg in having clear $J$-band H$_2$O absorption, and 
$H$-band and $K$-band CO absorption. The H$_2$O is weaker in \pg, 
while the CO is stronger.
We note that the CO in V1057 Cyg is atypically weak for FU Ori stars \citep{connelley2018}.


In terms of atomic lines, weak \ion{Na}{1}, 
\ion{Mg}{1}, \ion{Al}{1}, \ion{Si}{1}, and \ion{Ca}{1}
absorption are all present.
There is also weak \ion{H}{1} Pa$\beta$ absorption at $1.28~\mu$m.
The metal line strengths are similar to those seen in the comparison object 
Gaia 17bpi,  and somewhat stronger than in V1057 Cyg.
\pg\ and Gaia 17bpi also have comparable Pa$\beta$, 
which is likely blueshifted and formed in a wind. 

Notably, the \ion{He}{1} 10830 \AA\ line in \pg\ shows
a clear P Cygni profile (see inset panel within Figure~\ref{fig:nires}). 
This is evidence of a strong wind or outflow,
with blueshifted absorption in \ion{He}{1} 10830 \AA\
frequently seen in FU Ori stars \citep{connelley2018}.
We estimate a maximum velocity around -450 kms$^{-1}$.

In the nomenclature of \cite{connelley2018}, \pg\ appears to be a bona fide
FU Ori star.  In addition to having had its eruption observed,
it satisfies all of the spectroscopic criteria:
CO absorption, H$_2$O absorption, VO or TiO absorption (weakly),
Pa$\beta$ absorption, lack of emission lines, weak metal absorption,
and \ion{He}{1} 10830 \AA\ absorption. 
The hydrogen and helium lines likely have some outflow aspect 
to their absorption.  The metal lines originate in the source photosphere, 
however, and combined with the molecular line absorption, 
indicate a mixed temperature spectrum.

\section{Discussion}

The dramatic brightening behavior of \pg\ is detected most obviously 
in the mid-infrared (Figure~\ref{fig:midir}). 
The combined Spitzer, WISE, and NEOWISE data provide a long duration lightcurve.  
As the source has brightened,
it has become detectable in the near-infrared as well (Figure~\ref{fig:gattini}).
In the \pgitself\ data, which is higher cadence, the brightness of \pg\ is currently 
near the survey limit due to the high background in this field, 
and the lightcurve is thus somewhat noisy, though well-sampled.  
Over only the few months prior to manuscript submission, 
\pg\ has appeared near the faint limit 
in the ZTF alert stream, sampling the red optical. 
Given the evidence for a current (second) plateau phase, 
the source may not become amenable to study in the optical, however. 

A two-step rise is indicated by the mid-infrared lightcurve,  
each step being preceded by a shallow rise phase.
The earliest available data for \pg\ show a shallow brightening
between 2004 and 2009. Then
\pg\ exhibited a rapid (-0.25 mag/month) blue brightening 
by 1.8 mag in $W1$ and 0.8 mag $W2$ 
between 2009 September 6 and 2010 January 26. 
This was followed by a shallower slope (-0.01 mag/month), 
long-term colorless brightening over 2010-2018, 
amounting to only another 0.5 mag in both W1 and W2. 
The slope of the brightening increased during the latter part of 2018,
and switched in 2019 to a much steeper rise rate (-0.12 mag/month).  
At this point, the source became detectable by \pgitself\ in the near-infrared.
\pgitself\ (1.2 $\mu$m) and NEOWISE (3.4 and 4.6 $\mu$m) both record 
the same rise slope in this time frame,
and both show a transition to a plateau during 2020.
The total brightening of \pg\ over the past 16-25 years is $\sim5-5.5$ mag. 

Putting these lightcurve results into context, the 
recent rise slope compares well to the typical rise rates that can be derived 
for the optical lightcurves of historically documented FU Ori outbursts. These have a wide range, 
but are most frequently about $-0.05$ to $-0.2$ mag/month (Hillenbrand, 2021, in preparation).
Regarding the two-step rise in the mid-infrared lightcurve, a similar phenomenon
was recorded for the confirmed FU Ori source Gaia 17bpi \citep{hillenbrand2018}.
However, in that object the total duration of the first plateau phase was shorter, 
only 1-2 years, instead of the 5-10 years measured here for \pg.

We can estimate a source luminosity near the outburst peak 
based on the brightness level of the apparent plateau in the lightcurve 
($J\approx 14.6$ mag from \pgitself\ and $r\approx 20.5$ mag from ZTF),
and the adopted distance $d=2.81$ kpc.  The bolometric correction is of course unknown, 
but adopting values appropriate for a mid-F type star leads to a luminosity estimate
of 10-12 $L_\odot$, with perhaps a factor of 3 uncertainty if the bolometric correction is $\pm 1$ mag.
This is on the far low side of FU Ori outburst peak luminosities. Only
Gaia 17bpi, the only other source known to have exhibited a two-step 
mid-infrared brightening, like \pg\ has, and HOPS 383, an embedded young star
outburster whose status as an FU Ori type outburst is unclear, have luminosities this low.

Concluding our discussion, we note that
the best studied FU Ori stars are the optically visible portion of the population,
especially those bright enough for high dispersion spectroscopy.
However, a sizable fraction of the known FU Ori stars, about 1/3, 
are seen only at infrared wavelengths due to high extinction 
\citep[e.g. the $A_V = 20-50$ mag population in ][]{connelley2018}.  
Indeed, theoretical predictions are that earlier stage young stellar objects
experience a higher rate of FU Ori type outbursts \citep{bae2014}. 
Most of the known optically invisible group have been designated as FU Ori objects 
through source brightening at K-band that was noticed only post facto,
rather than while it was actually happening, along with spectroscopic confirmation.  
Embedded FU Ori sources thus generally do not have quiescence-to-rise lightcurves available.

Because it meets all applicable criteria for the FU Ori designation,
notably the spectral absorption signatures, lack of emission lines other than
P Cygni structure, and the total lightcurve amplitude,
\pg\ can be added to the list of optically invisible FU Ori candidates.
Unlike the other sources 
\footnote{We note that the VVV survey has produced some interesting 
candidates as well, recently \cite{guo2020}.},
\pg\ has had its photometric rise captured through 
wide-field continuous photometric monitoring programs.

\section{Summary}
We have discovered an outbursting young stellar object 
first detected at near-infrared wavelengths, but also seen 
in long duration mid-infrared time series data.
\pg\ is associated with the NGC 281-W molecular core,
a known area of recent and ongoing dense star formation activity.
Its salient features are:

\begin{itemize}
\item A Class I spectral energy distribution.

\item Foreground interstellar, molecular core, and circumstellar extinction
amounting to $A_V\approx 20$ mag.

\item A photometric rise that was detected in the 
\pgitself\ time-domain survey, and emphasized by comparing 
the recent $J$-band photometry 
with the lack of a $J$-band detection at the same position
more than two decades ago in the 2MASS all-sky photometric survey
or in the pointed work of \cite{megeath1997}, implying 
a total near-infrared brightness increase of $>5$ mag.

\item Extended nebulosity in our new $Ks$-band imaging, 
that was not present in 1990's images at the same wavelength,
with a cometary type structure $\sim14,000$ AU in size.

\item A measured 10-15 year photometric brightening trend exhibited in 
$3-5~\mu$m Spitzer $+$ WISE $+$ NEOWISE data, accumulating  
a total mid-infrared brightness increase of $>5$ mag.

\item Steepened mid-infrared brightening during 2019 that was
coincident with source detection at J-band.

\item Consistent rise rates during 2018-2020 in the 
mid-infrared NEOWISE lightcurve, and 2019-2020
in the near-infrared \pgitself\ lightcurve, amounting to $-0.12$ mag/month.

\item A flattening of the lightcurve from its rise phase to a plateau in 2020. 
The peak luminosity is estimated at a modest 10-12 $L_\odot$.

\item A near-infrared absorption line spectrum showing hydrogen line, metal line,
and molecular band absorption, indicative of a ``mixed temperature" spectrum.

\item Outflow as evidenced by a \ion{He}{1} 10830 \AA\ profile, 
with blueshifted absorption seen to about -450 kms$^{-1}$.
\end{itemize}

We conclude that \pg\ is a bona fide FU Ori star. 
The object warrants further study, 
especially at high spectral dispersion, 
as its outburst continues to develop.

\acknowledgements
Palomar Gattini-IR (PGIR) is generously funded by Caltech, Australian National University, the Mt Cuba Foundation, the Heising Simons Foundation, the Bi- national Science Foundation. PGIR is a collaborative project among Caltech, Australian National University, University of New South Wales, Columbia University and the Weizmann Institute of Science. MMK acknowledges generous support from the David and Lucille Packard Foundation. 

\facilities{Palomar Gattini IR, PO:1.2m:ZTF, PO:1.5m:SEDM, Hale:TSPEC, Keck/NIRES, Keck/DEIMOS, Spitzer, WISE, NEOWISE, IRSA}


\end{document}